# Relation between Superconducting Gap Minima and Nesting Vector in YNi$_2$B$_2$C

T. Baba,[1,*] T. Yokoya,[1,2] S. Tsuda,[1,†] T. Watanabe,[1,‡] M. Nohara,[3,§] H. Takagi,[3] T. Oguchi,[4] and S. Shin[1,5]

[1]*Institute for Solid State Physics, University of Tokyo, 5-1-5 Kashiwanoha, Kashiwa, Chiba 277-8581, Japan*

[2]*The Graduate School of Natural Science and Technology, Okayama University, 3-1-1 Tsushima-naka, Okayama 700-8530, Japan*

[3]*Department of Advanced Materials Science, University of Tokyo, 5-1-5 Kashiwanoha, Kashiwa, Chiba 277-8565, Japan*

[4]*Department of Quantum Matter, Graduate school of Advanced Sciences of Matter (ADSM), Hiroshima University, 1-3-1 Kagamiyama, Higashi-Hiroshima 739-8526, Japan*

[5]*RIKEN SPring-8 Center, 1-1-1 Koto, Sayo-chou, Sayo-gun, Hyogo 679-5143, Japan*

## Abstract

We have performed ultrahigh-resolution angle-resolved photoemission spectroscopy to directly observe the large superconducting (SC) gap anisotropy (GA) of YNi$_2$B$_2$C. The result shows large SC GA with a smooth variation along an intersection of a Fermi surface (FS) around the Γ-Z line and nearly isotropic SC gap values for two intersections of FSs around the X-P line but with a point-like non-zero minimum only for one sheet. The point-like SC gap minimum can be connected by the nesting vector reported from band calculations. The results show unexpectedly complicated SC GA of borocarbide superconductors.

74.25.Jb, 74.70.Dd, 79.60.-i

In the BCS (Bardeen, Cooper, Schrieffer) theory, superconductivity is caused by electron-electron pairing with momentum ($k$)-independent electron-phonon interaction, leading to an isotropic $s$-wave superconducting (SC) gap [1]. However, it is believed that this is not the case for high-$T_c$ cuprates and superconductors of heavy fermion, ruthenate, and organic materials, which are known to have highly anisotropic SC gaps with zero-gap regions (nodes) reflecting the change of sign in the pairing wave function [2]. In such materials, the pairing mechanisms other than phonon are actively discussed. On the other hand, regarded as phonon-mediated superconductors from band structure calculations (BCs) [3-5] and isotope effect studies [6,7], borocarbide superconductors $YNi_2B_2C$ and $LuNi_2B_2C$ [8,9] have turned out to have large SC gap anisotropy (GA) from various experimental studies [10-14].

Unprecedentedly large SC GA in a phonon-mediated superconductor has motivated various direction-dependent studies in order to determine the direction and type of nodal structure. The magnetic field orientation dependence of thermal conductivity [15], ultrasonic attenuation for all the symmetrically independent elastic modes [16], and scanning tunneling microscopy/spectroscopy in the vortex state [17] have reported point nodes located along [100] and [010] directions. On the other hand, nearly the same oscillation amplitudes at polar angle $\theta = 0°$ and 26° observed in field-angle-dependent heat capacity suggested a line-like nodal structure [18]. Also, isotropic two-gap superconductivity has been proposed by specific heat [19] and directional point-contact spectroscopy measurements [20]. Thus, the SC-gap structure is still controversial. Moreover, while those studies reported the direction of the nodes, the position of the nodes on Fermi surface (FS) sheets or FS-sheet dependence of SC gap, which is essential for considering the origin of the large SC GA, has not been clarified yet. Therefore, experimental determination of the type of nodal structure and the $k$-dependence of SC gaps is highly desired. For this purpose, high-resolution angle-resolved photoemission spectroscopy (ARPES) is a powerful experimental technique that can directly observe temperature ($T$)-dependent $k$-resolved electronic structure of solids, and thus a suitable technique to study the SC-gap structure and location of the nodes on FS sheets.

In this letter, we report ARPES results of (001) single crystal $YNi_2B_2C$ observing experimental valence-band dispersions, FSs, and $k$-dependent SC gaps. Experimentally determined FS sheets are found to be in good agreement with BCs. From $k$-dependent SC-gap measurements over several FS sheets, we found the $k$-dependence and FS-sheet dependence of SC gap. We also found that a cylindrical FS sheet around X-P shows the smallest SC-gap value, suggestive of a point-like nodal structure. More importantly, this SC gap minimum is found to be connected by the known nesting vector. These results indicate that nested FS plays a crucial role for the highly anisotropic SC gap.

Single crystals of $YNi_2B_2C$ were grown by a floating zone method [21]. SC transition $T$ ($T_c$) of 15.4 K was determined by the onset of magnetic susceptibility measurements. The residual resistivity ratio estimated from resistivity measurements was 37.4. Sample orientations were measured *ex situ* by using Laue backscattering and further confirmed *in situ* by the symmetry of ARPES spectra. All the ARPES data presented here have been measured with a high-resolution hemispherical analyzer using the monochromatic He Iα resonance line (21.218 eV). Energy resolutions of valence-band and SC-gap measurements were set to ~ 30 meV and 1.7 meV, respectively. The angular resolution was set to 0.18°, corresponding to $k$ resolution of ± 0.0067 Å$^{-1}$. The sample $T$ for valence-band measurements was 20 K, while those of SC-gap measurements were 6 K (SC state) and 18K (normal state), respectively. The base pressure of our spectrometer was better than $2 \times 10^{-11}$ Torr. Single crystal clean surfaces of $YNi_2B_2C$ (001) were prepared *in situ* by repeating Ar-ion bomberdment for several hours and flash heating at > 1200 °C. No traces of oxygen, sulfur nor other impurities were detected in the Auger electron (AE) spectra (Fig.1 (a)), and a clear (1 × 1) low energy electron diffraction (LEED) pattern (inset of Fig.1 (a)) having the same symmetry as the bulk crystal was observed. The Fermi level ($E_F$) of samples was referenced to that of a gold film evaporated onto the sample substrate and its accuracy was estimated to be better than 5 meV for valence-band measurements and 0.2 meV for SC-gap measurements. The BCs of $YNi_2B_2C$ were performed by a full potential linearized augmented plane wave (FLAPW) method for detailed comparison with FS sheets observed experimentally. ARPES defines the $k_{//}$ component of an electron. For three-dimensional electronic structures of $YNi_2B_2C$, perpendicular component $k_\perp$ can be obtained by assuming a free electron final state model with an inner potential $V_0$. Here we used an inner potential $V_0$ = 13.5 eV by comparing observed FS sheets with calculated ones.

Figure 1 (c) shows ARPES spectra from $YNi_2B_2C$ (001) measured as a function of the Polar angle. Corresponding measured $k$ points are on the red curve in Fig. 1 (b) along the Γ (Z)-X (P) high symmetry direction in the BZ. Several pronounced peaks, whose intensity and energy position systematically change as a function of the measured $k$, are indeed observed, indicative of band dispersions of $YNi_2B_2C$ (001). In Fig. 1 (d), we show an ARPES intensity map as functions of binding energy and *k*. White regions, which have higher intensity, correspond to electronic bands. We observed several dispersive bands especially within 4 eV of $E_F$ and found some of them approach toward $E_F$. To observe FS sheets, we plotted ARPES intensities of a 12.5 meV window centered at $E_F$ over the BZ, as shown in Fig. 2 (a), where white regions correspond to FS sheets. We observed two high intensity lines perpendicular to the Γ (Z)-A line. We also found that a high intensity region around the X (P) point that can be divided into two square-like sheets centered at the X (P) point due to a lower intensity gap (see also Fig. 4). In Fig. 2 (b)-(d), we also show three FS sheets (from 17 th, 18 th, and 19 th bands, respectively) calculated by the FLAPW method. To correlate observed intensity variation with calculated FS sheets, we superimposed intersections of calculated FS sheets with a measured *k* plane upon experimental FSs. The two high intensity lines perpendicular to Γ (Z)-A can be ascribed to the two walls of an elliptic intersection of the FS sheet of 17 th band. The two components of the higher intensity region around X (P) can be ascribed to FS sheets of 17 th and 18 th bands.

The observed intensity variation at $E_F$ being able to be correlated with calculated FS sheets, we have performed low-*T* ultrahigh-resolution ARPES for SC and normal states in steps of 1° for all points of 1st BZ to investigate the *k*-dependence of SC gaps, as shown in Fig. 3. Figure 3 (a) shows regions of FS sheets for which we measured SC gap values and Fig. 3 (b) shows ARPES spectra for selected points on FS sheets. We find that all ARPES spectra (including spectra not shown in this manuscript) exhibit a remarkable *T*-dependence, indicative of opening of a SC gap below $T_c$. Sizable SC gap opening at every measured *k* point of 1st FS suggests absence of a line node along the $k_Z$ direction within measured FS sheets.

To investigate the *k*-dependence of magnitude of SC gaps (Δ), we numerically fitted the data with the modified BCS function [22], as described previously for the $MgB_2$ [23]. The modified BCS function was first multiplied with a Fermi-Dirac (FD) function of the measured *T* and then convoluted with a Gaussian with a full width at half maximum equal to the experimental energy resolution. Δ values for all measured points along FSs are plotted in Fig.3 (c). One can observe larger SC gap values of the elliptic intersection of the 17th FS sheet than those of X (P)-centered 17th and 18th FS sheets. As for the *k*-dependence, the SC gap values from the both sides of the elliptic intersection of the 17th FS sheet show marked *k*-dependence with minimum along the Γ (Z)-A direction and systematic increase away from it. Contrary, the SC gap values along the intersections of X (P)-centered 17th and 18th FS sheets are found nearly isotropic but with a point-like minimum along the $\Gamma$ (Z)-X (P) direction only on the 17th FS sheet. This point-like minimum has the lowest SC gap value of 1.5 meV within the measured $k$ plane. This result indicates that the SC-gap structure is not an isotropic two-gaps suggested by Ref [19,20]. The result also shows that the nodal structure does not possess a line node at least on 17th and X (P)-centered 18th FSs, which are expected to have dominant contribution to the total density of state (DOS) and therefore to the SC GA observed from other experiments.

In borocarbide superconductors, it is suggested that the FS has a nesting feature whose nesting vector is Q ~ (0.55, 0, 0) from BCs [24], which was experimentally suggested by two-dimensional angular correlation of electron-positron annihilation [25]. In addition, from inelastic neutron scattering measurements [26], pronounced Kohn anomalies have been observed with the significant phonon softening at wave vector close to Q ~ (0.55, 0, 0). Moreover, nuclear magnetic resonance study in the normal state [27,28], $1/T_1T$ of $^{11}$B increases monotonously as $T$ decreases though the Knight shift which correspond to the susceptibility at $q = 0$ weakly depends on $T$. The results suggested an enhancement of antiferromagnetic (AF) fluctuations at low $T$, which are expected to originate from the nesting of the FSs because the generalized susceptibility from BCs shows a peak at $q \sim Q$ [24].

In order to understand the relationship between the observed SC gap minima and the nesting vector, we plotted the ARPES intensity map over the BZ by using the four-fold symmetry of the crystal, as shown in Fig. 4. Corresponding SC gap minima observed on X (P)-centered and $\Gamma$ (Z)-centered 17 th FS (**D** and **A** in Fig. 3) are connected by vectors I and II, respectively. Estimated magnitudes of wave vectors are $0.60 \pm 0.03$ for the vector I and $0.56 \pm 0.04$ for the vector II. As for the vector I, it has the same direction and nearly the same magnitude as the nesting vector from previous works [24,25]. More importantly, location of the vector I on the observed FSs is also in excellent agreement with the previous works [24,25]. A slight enlargement of the magnitude may be due to the difference in $k_z$ ($k_z \sim 0.25$ for the vector I, but $k_z = 0$ of the known nesting vector), which is consistent with the observation that the experimental FS sheets do not show a "crater structure" that is expected to be located around $k_z = 0.0$ from BCs [24]. A dip structure in the ARPES spectrum located at ~ 6 meV (indicated by an arrow in Fig.3 (b) **D**) which correspond to the sum of two energies of SC gap (1.5 meV) and soft phonon (~ 4 meV) [26] may give another evidence of the nesting induced reduction of SC gap. On the other hand, while the vector II has the same direction and magnitude as the known nesting vector [24,25], its location on the FS is different from the previous works [24,25].

More recently, a quasiclassical Eilenberger theory taking into the two-dimensional FS calculated by local density approximation concluded that point nodes are located at parts connected by the known nesting vector at $k_z = 0.0$ [29]. They also showed local non-zero SC gap minima at $k_z = 0.5$. Thus the ARPES experimental evidence showing sudden decrease of SC gap at point D connected with the vector I agrees well with this theoretical study. The location on FSs and $k_z$ (~ 0.375) of the vector II can be related to the vector B in Ref. [29] that connects two calculated $k_F$'s at $k_z = 0.5$ with the AF fluctuations having (0.55, 0, 0), though Ref. [29] does not expect any local SC gap minima on the two calculated $k_F$'s at $k_z = 0.5$. Recently, Kontani proposed a theory that microscopically explained a mechanism of an anisotropic $s$-wave superconductivity with deep SC gap minima by solving the strong-coupling Eliashberg equation in a system where strong electron-phonon coupling and moderate AF fluctuations coexist under an assumption of isotropic FS [30]. According to this theory, each pair of SC gap minima is connected by the nesting vector. This is in line with present observations.

In conclusion, we have successfully studied the $k$-dependent SC gap of YNi$_2$B$_2$C (001) by using low-$T$ ultrahigh-resolution ARPES. The results provide direct experimental evidence for the anisotropic SC gap, which is in line with the point node as the SC-gap structure. The $k$ positions of one of the local SC gap minima can be correlated with the known nesting vector, suggesting close correlation between nodes and the nesting vector. On the other hand, another SC gap minimum has also been observed, which can not be related to the known nesting vector but can be correlated with the pairs of $k_F$ with known AF fluctuations. The present study, thus, reveals very complicated SC gap anisotropy of a borocarbide superconductor, which leads to deeper understanding for anisotropic superconductivity.

We thank Profs. H. Harima, K. Machida, and Y. Ichioka for very valuable discussion. We also thank Dr. T. Aizawa for showing their system for preparing clean single crystal surfaces. This work was supported by Grant-in-aid from the Ministry of Education, Science, and culture of Japan and Culture of Japan.

References

* Present address: Research Department, NISSAN ARC, Limited, Yokosuka, Kanagawa 237-0061, Japan

Electronic address: t-baba@nissan-arc.co.jp

[†] Present address: National Institute for Material Research, Tsukuba, Ibaraki 305-0047, Japan

[‡] Present address: Department of Physics, College of Science and Technology, Nihon University, Chiyoda-ku, Tokyo 101-8308, Japan

[§] Present address: Department of Physics, Okayama University, 3-1-1 Tsushima-naka, Okayama 700-8530, Japan

Figure captions

FIG. 1: (Color online) (a) first derivative AE spectrum of $YNi_2B_2C$ (001). Inset: the LEED pattern of $YNi_2B_2C$ (001)-1 × 1 at 64 eV electron beam energy. (b) Typical spherical path through the BZ at 21.218 eV (He Iα) for states at $E_F$. Upper right corner: BZ and the Γ-X plane. (c) ARPES spectra of $YNi_2B_2C$ (001) obtained with He Iα measured along the Γ (Z)-X (P) high symmetry direction. (d) The valence band intensity map of $YNi_2B_2C$ (001).

FIG. 2: (Color online) (a) The intensity map at $E_F$ (energy window = $E_F$ ± 12.5 meV) of $YNi_2B_2C$ (001), along with the calculated FSs. (b)-(d) calculated FSs from 17 th, 18 th, and 19 th bands, respectively.

FIG. 3: (Color online) (a) Regions of FSs for which we measured SC gaps. (b) The selected low-*T* high-resolution ARPES spectra. Blue and red circles represent the experimental results measured at 6 K (SC) and 18 K (normal), respectively. Green and yellow lines represent the fitting results by using the modified BCS function and the FD function, respectively. (c) A plot of SC gap values on those FSs, where the color of symbols corresponds to the color of measured *k* regions shown in Fig. 3(a).

FIG. 4: (Color online) Symmetrized FS. The arrow I and II represent the vector connected to the SC gap minima in X (P)-centered 17th FS and $\Gamma$ (Z)-centered 17th FS, respectively. Estimated wave vectors are $0.60 \pm 0.03$ for I and $0.56 \pm 0.04$ for II, receptivity.

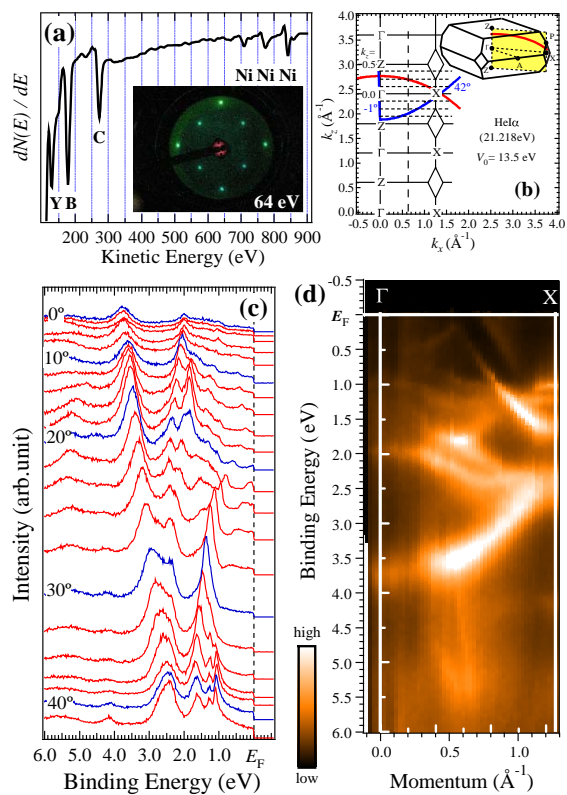

Figure 1 (T. Baba et al.)

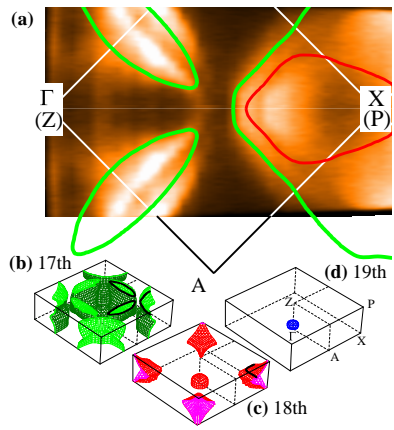

Figure 2 (T. Baba et al.)

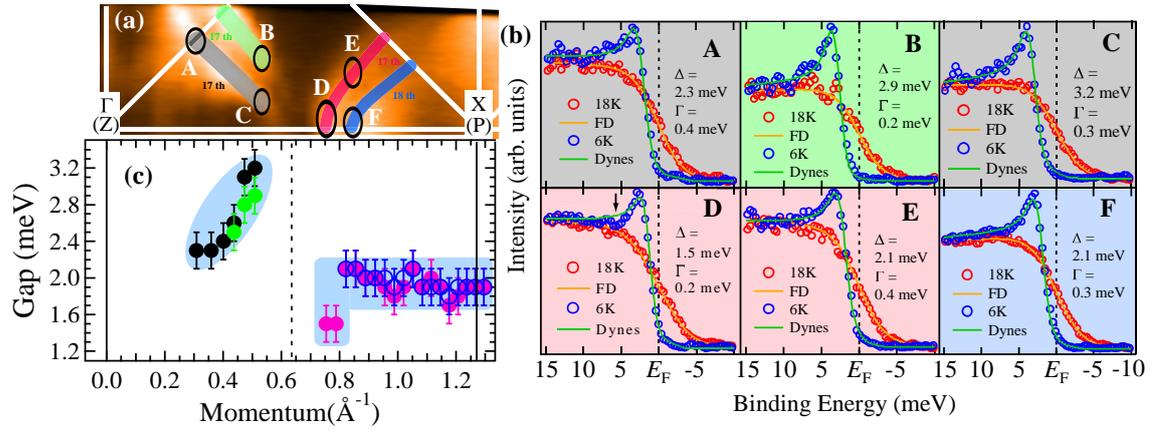

Figure 3 (T. Baba et al.)

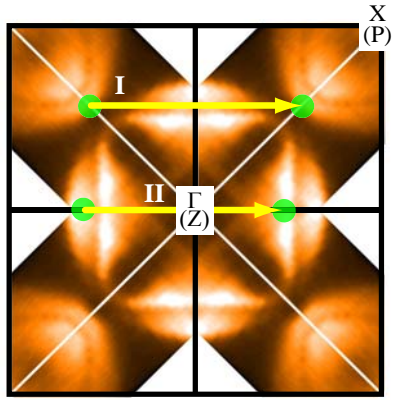

Figure 4 (T. Baba et al.)